\documentclass{ar-1col}

\usepackage[numbers]{natbib}
\usepackage{amssymb}
\usepackage{amsmath}

\def \k {\mathbf{k}}

\def \r {\mathbf{r}}
\def \x {\mathbf{x}}
\def \q {\mathbf{q}}
\def \d{\partial}
\def \Q {\mathbf{Q}}

\jname{Annual Review of Condensed Matter Physics}
\jvol{10:63-84}
\jyear{2019}
\doi{10.1146/annurev-conmatphys-031218-013339}

\begin{document}

% Page header
\markboth{Berg et al.}{QMC studies of critical metals}

% Title
\title{Monte Carlo Studies of Quantum Critical Metals}

%Authors, affiliations address.
\author{Erez Berg$^1$, Samuel Lederer$^2$, Yoni Schattner$^{3,4}$, and Simon Trebst$^5$
\affil{$^1$Department of Physics, James Franck Institute, University of Chicago, Chicago, USA 60637; email: berge@uchicago.edu}
\affil{$^2$Department of Physics, Massachusetts Institute of Technology, Cambridge, USA, 02139}
\affil{$^3$Department of Physics, Stanford University, Stanford, USA, 94305}
\affil{$^4$Stanford Institute for Materials and Energy Sciences, SLAC National Accelerator Laboratory and Stanford University, Menlo Park, CA 94025, USA}
\affil{$^5$Institute for Theoretical Physics, University of Cologne, Cologne, Germany, 50937}}

%Abstract
\begin{abstract}
%Abstract text, approximately 150 words.
Metallic quantum critical phenomena are believed to play a key role in many strongly correlated materials, including high temperature superconductors. Theoretically, the problem of quantum criticality in the presence of a Fermi surface has proven to be highly challenging. However, it has recently been realized that many models used to describe such systems are amenable to numerically exact solution by quantum Monte Carlo (QMC) techniques, without suffering from the fermion sign problem. In this article, we review the status of the understanding of metallic quantum criticality, and the recent progress made by QMC simulations. We focus on the cases of spin density wave and Ising nematic criticality. We describe the results obtained so far, and their implications for superconductivity, non-Fermi liquid behavior, and transport in the vicinity of metallic quantum critical points. Some of the outstanding puzzles and future directions are highlighted.
\end{abstract}

%Keywords, etc.
\begin{keywords}
quantum critical metal, superconductivity, non-Fermi liquid, Ising nematic order, spin density wave order, quantum Monte Carlo
\end{keywords}
\maketitle

%Table of Contents
\tableofcontents

% Heading 1
\section{Introduction}
One of the ubiquitous properties of strongly correlated materials is the presence of different electronically ordered states in close competition with each other, which can be tuned by relatively small changes in a tuning parameter such as pressure, material composition, or magnetic field. Another common feature of these systems is an anomalous metallic state, whose properties are incompatible with Fermi liquid theory. An appealing scenario to account for the latter observation is that it originates from an underlying {\it quantum critical point} (QCP) at which the metallic ground state becomes unstable towards some form of order. In this scenario, the behavior of the anomalous metallic state thus reflects its proximity to a non-Fermi liquid fixed point, which governs the system's properties over a broad range of temperature and tuning parameter.

%Evidence for QCP in materials

There is ample experimental evidence that some sort of quantum criticality is present in many materials, such as heavy fermion compounds~\cite{mathur1998magnetically,gegenwart2002magnetic,custers2003break,Paglione2003,bianchi2003avoided,Park2006,Gegenwart2008,nakatsuji2008superconductivity,matsumoto2011quantum,Landaeta2017}, iron-based superconductors~\cite{hashimoto2012sharp,Shibauchi2014,Chu2012,KouFisher2015,Gallais2013,Blumberg2014,Bohmer2014}, and electron-doped cuprate superconductors~\cite{motoyama2007spin,armitage2010progress}. In the hole-doped cuprates the situation is less clear, although there is some evidence for a QCP underneath the superconducting dome in these systems, as well~\cite{Daou-2008b,Ramshaw2015,tailleferunp}. In all these systems, a ``fan''-shaped metallic regime emanates from the (putative) QCP, often featuring, among other phenomena, anomalous power laws in the temperature ($T$) dependence of the electrical resistivity. (Rather than the $T^2$ behavior of Fermi liquid theory, the resistivity in the putative quantum critical regime is often linear in $T$\footnote{Other power laws are seen in certain systems. See, for example, Refs.~\cite{Gegenwart2008,nakatsuji2008superconductivity,wolfle2011quasiparticles}.}.) In almost all the the above systems, an unconventional superconducting phase emerges with a maximum transition temperature close to the apparent QCP, indicating an intimate connection between quantum criticality and superconductivity.

%Theory of metallic QCP

The theory of quantum criticality in metals is a decades-old problem, dating back to the seminal works of Hertz~\cite{Hertz1976}, Moriya~\cite{Moriya}, and Millis~\cite{Millis1993}.
%This problem turns out to be particularly challenging , since the low-energy degrees of
%freedom include both the fermionic quasiparticles and the bosonic fluctuations of the order
%parameter; the two sets of energy degrees of freedom are strongly coupled to each other, and have to be treated on
%equal footing.
The problem has proven challenging due to the profusion of low energy degrees of freedom.
Unlike classical critical phenomena and QCPs in insulators, where the correlations typically become singular only at a single point in momentum space, in the metallic case the low-energy degrees of freedom live on an entire extended manifold of momenta--the Fermi surface\footnote{The situation is different in semimetals, for instance systems with a Dirac dispersion, where the ``Fermi surface'' consists of only one point. In these systems, much theoretical progress has been made; see, e.g., Refs.~\cite{Vojta2000,Assaad2013,li2017fermion}.}. As we shall elaborate below, despite intense work on the topic and significant technical progress~\cite{altshuler1994low,nayak1994renormalization,nayak1994non,Altshuler1995,chakravarty1995,Castellani1995,Abanov1999,Abanov2000,Abanov2003,Metzner2003,Abanov2004,Kotliar2004,Lawler2006, rech2006quantum,lohneysen-2007,Aji2007,SSLee2009,Woelfle2009,Metlitski2010,Metlitski2010a,mross2010controlled,Maslov2010,Dalidovich2013,efetov2013pseudogap,abrahams2014strong,Fitzpatrick2014,meier2014cascade,Holder2015,Wang2015,Raghu2015,varma2015quantum,varma2016quantum,meszena2016nonperturbative,schlief2017exact,lee2017recent,saterskog2018framework,}, there is no accepted theory to describe quantum criticality in systems with a Fermi surface.
%Empirically, it is not even clear that such a ``universal theory'' exists; there seem to be different classes of QCPs with different qualitative behavior (e.g., different apparent critical exponents) in different systems between the {\it same} $T=0$ phases of matter~\cite{coleman2001fermi,si2001locally,Senthil2008}.

In recent years, it has been appreciated that many paradigmatic models of metallic criticality can be solved~\cite{Berg2012,Schattner2016,li2016makes,Schattner2016a,gerlach2017quantum,Lederer2017,wang2017charge,wang2017superconductivity,xu2017non,Li2017,gazit2017emergent} by quantum Monte Carlo (QMC) simulations without suffering from the notorious ``sign problem''~\cite{Loh1990},
which often hinders QMC simulations in systems of interacting fermions. This opens the way for {\em fully controlled} and {\em numerically exact} solutions of models of quantum critical metals, providing a non-perturbative handle on the problem. Beyond their quantitative guidance, these numerical solutions are useful both as a benchmark for existing theoretical descriptions, and as a guide for new ones. The purpose of this paper is to review the status of the theory of metallic quantum criticality, focusing on the recent progress made and questions raised by sign problem-free quantum Monte Carlo simulations, and their implications for our current understanding of this problem.

\section{Setup of the problem}
We start with a discussion of general theoretical considerations for the problem of metallic quantum criticality. We present a (presumably) generic field theoretical description of such quantum critical points, and review briefly the various analytical techniques applied to it, before turning to the numerical QMC approach in the next section.
\subsection{General considerations and field-theoretical model}

Consider a lattice model of fermions, coupled via a general-short range interaction. For a generic dispersion of the fermions (i.e. without the Fermi surface fine tuned to perfect nesting or a van Hove singularity), the model is stable when interactions are infinitesimally weak, with the exception of a possible superconducting instability. Thus, any quantum critical point at which an ordered state forms out of the metal must occur at {\em finite} strength of the coupling. From this perspective, the problem of metallic quantum criticality is intrinsically an intermediate coupling problem, and may be not accessible via methods that are perturbative in the interaction strength.

In order to enable a perturbative approach to the critical point, we may introduce a fluctuating order parameter field  by hand, tune it to the vicinity of a symmetry-breaking transition, and consider the effects of a small coupling between the order parameter fluctuations and the fermions. Schematically, the system is described by the following Euclidean action:
\begin{eqnarray}
S = S_{\psi} + S_{\phi} + S_{\mathrm{int}},
\label{eq:action}
\end{eqnarray}
where
\begin{eqnarray}
S_{\psi} &=& \int_0^\beta d\tau \, \sum_{\k} \psi_{\mathbf{k}}^\dagger \left( \partial_\tau + \varepsilon_{\mathbf{k}} - \mu \right) \psi_\mathbf{k}, \nonumber \\
S_{\phi} &=& \int_0^\beta d\tau \int d^d x \left[ \frac{1}{2} r \phi^2 + \frac{1}{2}(\nabla \phi)^2 + \frac{1}{2c^2} (\partial_\tau \phi)^2 + \frac{u}{4} \phi^4 + \dots \right].
\label{eq:S}
%S_{\mathrm{int}} &=& \lambda \int_0^\beta d\tau \int d^d x \, \phi(\mathbf{x}) \psi^\dagger (\mathbf{x}) \psi (\mathbf{x}).
\end{eqnarray}
Here, the fermionic action $S_{\psi}$ is given in terms of the  fermionic operators $\psi^\dagger, \psi$ (which in general may have spin and other indices, not displayed here) with dispersion $\varepsilon_{\mathbf{k}}$ and chemical potential $\mu$.
The bosonic action, $S_{\phi}$, is the usual Landau-Ginzburg-Wilson action, written as an expansion in powers of the order parameter $\phi$  and its derivatives. The precise nature of $\phi$ depends on the symmetry broken in the ordered state (see explicit examples below). The  tuning parameter $r$ drives the system through a QCP.
The interaction term, $S_{\mathrm{int}}$, is of the ``Yukawa'' form, linear in the order parameter and quadratic in the fermion operators, and includes a form factor that encodes the symmetry of the order parameter (see Eqs.~(\ref{eq:SDW},\ref{eq:nematic}) below). We will focus on the case of $d=2$ spatial dimensions, relevant to many materials of interest.

%Physically, the order parameter field $\phi$ can originate from integrating out high energy fermionic modes, generating analytic terms in $\phi$ and its gradients~\cite{Abanov2003}. Then, the action (\ref{eq:action}) is interpreted as the resulting low-energy effective theory, containing the remaining low-energy fermion modes and the long-wavelength order parameter fluctuations. Alternatively, the action (\ref{eq:action}) can emerge as a description of a system of itinerant electrons coupled to distinct bosonic degrees of freedom, such as localized spins or phonons near a structural phase transition. From the symmetry point of view, these two scenarios are equivalent, and therefore their universal aspects should be the same. The bare values of the parameters in $S_{\phi}$ can strongly depend on the physical origin of the order parameter; for example, if the field $\phi$ represents a displacement of a phonon mode, then the bare boson velocity $c$ is usually much smaller than the Fermi velocity.

\begin{figure}[t]
\includegraphics[width = \columnwidth]{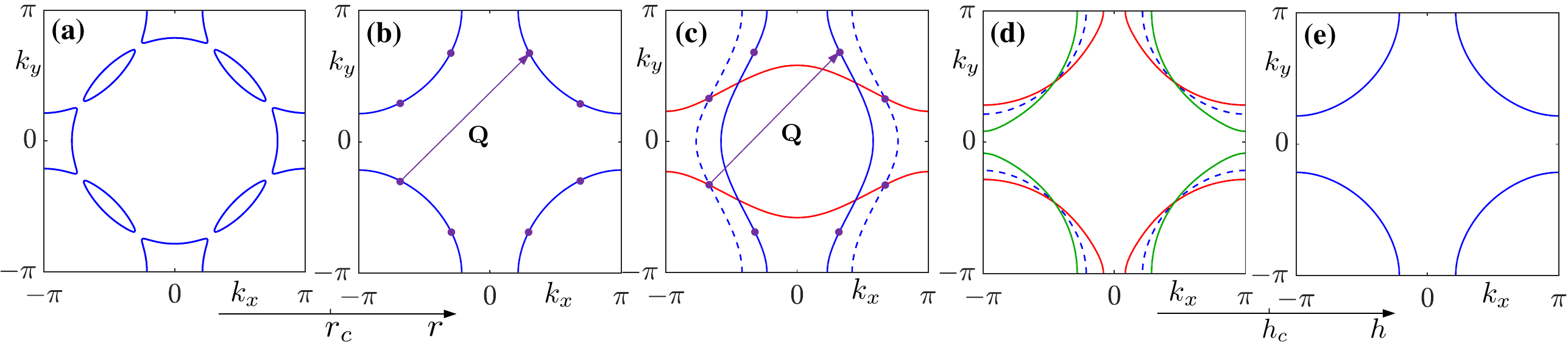}
\caption{Evolution of the Fermi surface across the SDW transition (a,b) and the Ising nematic transition (d,e). Panel (c) shows the band structure of the two-band model used in the QMC simulations of a SDW transition~\cite{Schattner2016a,gerlach2017quantum}. The dashed line is the blue Fermi surface shifted by $\Q = (\pi,\pi)$. In (b,c) the locations of the hot spots, connected by the magnetic ordering vector $\Q = (\pi,\pi)$, are indicated by the circles. }
\label{fig:FermiSurface}
\end{figure}

Throughout this paper, we will focus on two types of paradigmatic quantum critical points: a spin density wave (SDW) QCP and an Ising-nematic QCP. A SDW QCP involves ordering at a non-zero wavevector $\mathbf{Q}$. For simplicity, we focus on the case of a commensurate ordering wavevector (antiferromagnetic order). The coupling of the order parameter to the fermions has the form
\begin{equation}
S_{\mathrm{int}} = \lambda \int d^2 x \int_0^\beta d\tau\, e^{i \Q \cdot \x}\vec{\phi}(\x) \cdot \psi^\dagger(\x) \vec{\sigma} \psi(\x) + h.c.,\ (\mathrm{SDW})
\label{eq:SDW}
\end{equation}
where $\vec{\sigma}$ are spin Pauli matrices, and $\vec{\phi}$ is the SDW order parameter, which may be a $1$-, $2$-, or $3$-component vector (depending on whether the SDW order parameter has easy-axis, easy-plane, or isotropic character, respectively). The SDW order parameter couples particularly strongly to fermions in the vicinity of a discrete set of ``hot spots'' on the Fermi surface (or hot lines in $d=3$ dimensions), connected to each other by the magnetic ordering wavevector $\Q$. At these points, fermions can scatter off the low-energy order parameter fluctuations while remaining on the Fermi surface. In the SDW ordered phase, a gap opens at and near the hot spots, causing reconstruction of the Fermi surface, as shown in Fig.~\ref{fig:FermiSurface}(a,b).

A second type of QCP we will consider is an Ising-nematic QCP, where the point-group symmetry of the system is reduced from tetragonal to orthorhombic. This is a paradigmatic example of a $\Q=0$ order parameter which preserves translational symmetry. The order parameter has two inequivalent configurations, and is represented by an ``Ising-like'' scalar field that changes sign under rotation by $90^\circ$. The interaction term in the action reads
\begin{equation}
S_{\mathrm{int}}= \lambda \int d^2 x \int_0^\beta d\tau\, {\phi}(\x) \psi^\dagger(\x) (\d_x^2 - \d_y^2) \psi(\x)\,\ (\mathrm{Ising\,\,nematic}).
\label{eq:nematic}
\end{equation}
In the ordered phase, $\phi\ne 0$, and the Fermi surface becomes elongated either along $x$ or $y$ [see Fig.~\ref{fig:FermiSurface}(d,e)]. In the vicinity of the QCP, the small--$\Q$ fluctuations of the order parameter couple strongly to fermions along the entire Fermi surface (with the exception of a discrete set of ``cold spots'' along the diagonals, where the coupling vanishes to leading order in $\phi$).

\subsection{Perturbative expansions}
\label{sec:perturbative}
In all cases, the
%for $\lambda=0$, the QCP is in the the Wilson-Fisher $d=2+1$ universality class. The
coupling to the fermions $\lambda$ is a {\it relevant} parameter, in the sense that perturbation theory in $\lambda$ breaks down  at sufficiently low energy scales.
%, driving the system away from this fixed point.
It is nevertheless useful to start from the small $\lambda$ limit, which can provide information about the crossover regime and give clues about the nature of the ground state.

Let us focus on the case of an Ising-nematic QCP, which is somewhat simpler to analyze. To leading order in $\lambda$, the bosonic field acquires a self-energy given by\footnote{Here, we set the units such that $\lambda^2$ has dimensions of energy.}
\begin{equation}
\Pi(\q,i\Omega_n) = -\lambda^2 \cos^2(2\theta_\q) \nu_0 \left( 1 - \frac{|\Omega_n|}{\sqrt{\Omega_n^2 + (v_F^2 q^2)}} \right).
\label{eq:Pi}
\end{equation}
Here, $\nu_0$ and $v_F$ are, respectively, the density of states of the fermions at the Fermi level and their Fermi velocity, and $\theta_\q$ is the angle between $\q$ and the $x$ axis. For simplicity, we have assumed a circular, isotropic Fermi surface. The angular dependence of (\ref{eq:Pi}) comes from the anisotropy of the Yukawa coupling (\ref{eq:nematic}). In particular, note that (\ref{eq:Pi}) vanishes along the diagonals $q_x = \pm q_y$; this is a consequence of the ``cold spots'' where the coupling constant between $\phi$ and the fermions vanishes.

%Eq.~(\ref{eq:Pi}) has immediate consequences for the character of the low-energy dynamics of the bosonic critical fluctuations. Originally, the bosonic part of the action~(\ref{eq:S}) is characterized by a dynamical critical exponent $z=1$ (i.e., the characteristic frequencies and momenta scale in the same way).

At sufficiently low frequency, the $|\Omega_n|$ term in Eq.~(\ref{eq:Pi}) becomes dominant over the $\Omega_n^2/c^2$ term in Eq.~(\ref{eq:S}). The crossover scale where this occurs can be roughly estimated by equating the frequency-dependent part of (\ref{eq:Pi}) to the bare inverse $\phi$ propagator, assuming that $|\q| \sim \Omega_n/c$. The resulting crossover scale is
\begin{equation}
\Omega_{b} = |\lambda| \sqrt{\nu_0 c^3/v_F}.
\end{equation}
At frequencies below $\Omega_b$, the boson dynamics is dominated by the ``Landau damping'' term proportional to $|\Omega_n| / (v_F q)$, and the dynamical critical exponent $z$ increases from $1$ to $3$.

Next, we consider the lowest-order contribution of the boson on the fermion self-energy. We focus on frequencies $\omega_n \ll \Omega_b$, where we should use the ``dressed'' form of the boson propagator, including its self-energy (\ref{eq:Pi}).\footnote{The existence of a well-defined intermediate regime where the bosons are strongly dressed but the fermions are only weakly renormalized is formally justified if we take the limit of large $N$, where $N$ is the number of fermion flavors.}
%\comment{SL: I only really understand the existence of this regime in large N. Do we want to stake this out here?}. %; we will check the self-consistency of this approach below.
To order $\lambda^2$, the self-energy on the Fermi surface is given by
\begin{equation}
\Sigma(\k_F, \omega_n) \sim - i (\lambda^4 \sin^4{\theta_{\k_F}} / E_F)^{1/3} |\omega_n|^{2/3} \mathrm{sgn}(\omega_n) \,,
\label{eq:Sigma}
\end{equation}
where $E_F \sim \nu_0 v_F^2$ is the Fermi energy, and $\theta_{\k_F}$ is the angle between the fermion's momentum and the $x$ axis. The fermions thus become strongly damped by the coupling to the bosonic fluctuations almost everywhere on the Fermi surface (with the exception of the ``cold spots'' where $\theta_{\k_F} = \pm \pi/4$), with a damping rate that scales as $\omega^{2/3}$ (compared to $\omega^2 \log(1/\omega)$ in an ordinary two-dimensional Fermi liquid).

From (\ref{eq:Sigma}), we can estimate the scale at which Fermi liquid behavior breaks down. As for the bosons, this is done by equating the self-energy (\ref{eq:Sigma}) to the $i\omega_n$ term in the bare fermion inverse propagator. This procedure gives
\begin{equation}
\Omega_{\mathrm{NFL}} = \frac{\lambda^4}{E_F}.
\label{eq:NFL}
\end{equation}
Note that for small $\lambda$, $\Omega_{\mathrm{NFL}} \ll \Omega_b$. Thus, there is a parametrically broad window of energies between $\Omega_{\mathrm{NFL}}$ and $\Omega_b$ where the bosons are described by overdamped dynamics (with a dynamical critical exponent $z=3$), but the feedback of the bosons on the fermions is weak. This justifies the use of the zeroth-order fermionic polarization bubble in the evaluation of Eq.~(\ref{eq:Pi}).

At frequencies below $\Omega_{\mathrm{NFL}}$, one can no longer ignore the fermion self-energy. Higher-order corrections to $\Sigma$ are found to be increasingly singular at low frequencies. %, and a non-perturbative solution is required.
%One way to limit the number of diagrams is to generalize the problem to the case of $N$ identical fermion flavors, all coupled to the same bosonic field, and perform a $1/N$ expansion. However,
%It turns out that
Extending the problem to the case of a large number $N$ of fermion flavors does not solve this problem, since
%Introducing a large number of fermionic flavors
a large set of diagrams that are naively subleading in powers of $1/N$ are in fact divergent, and must be treated on equal footing
%upon proper regularization (performed by using dressed Green's functions in the internal fermionic  lines), all diagrams with a planar topology are found to contribute equally
~\cite{SSLee2009,Metlitski2010}. There is currently no understanding of the resulting fixed point.
%A heroic loop expansion up to four loop order~\cite{Holder2015} reveals that both $z$ and the anomalous dimension of the fermions are renormalized; however, there is no formal justification for stopping after fourth order.
The $1/N$ expansion similarly fails in the SDW QCP problem~\cite{Metlitski2010a}.

Other modifications of the problem have been devised to gain control of calculations~\cite{mross2010controlled,Dalidovich2013,Raghu2015,Fitpatrick2015}. These methods, which we do not review here, involve various large-$N$ limits, the introduction of non-local terms in the action, and the extension of the problem to fractional dimensions. The properties of the $d=2$, finite-$N$ problem remain unclear.
%One approach~\cite{mross2010controlled} is to replace the $q^2$ term in $S_\phi$ [Eq.~(\ref{eq:S})] by $|\q|^{1+\varepsilon}$, and perform a double expansion in $\varepsilon$ and $1/N$. To leading order in $\varepsilon$, the $z=3$ fixed point, with the boson and fermion self-energies behaving as (\ref{eq:Pi}) and (\ref{eq:Sigma}), respectively, is found to be stable. However, this method has the disadvantage that the bosonic action is non-local in space for any non-integer $\varepsilon$; it is not clear to what extent the properties of the fixed point are artifacts of this non-locality. An alternative approach is to expand in the spatial dimension, keeping the dimension of the Fermi surface fixed at $1$~\cite{Dalidovich2013}; in this approach, the bosonic action is local, but it is not clear to what extent the result can be extended to spatial dimension $d=2$. Alternative expansion methods have been proposed~\cite{Raghu2015,Fitpatrick2015}, but the properties of the $d=2$, finite-$N$ problem remain unclear.

For the SDW problem, a strongly coupled fixed point was found recently~\cite{schlief2017exact,lunts2017emergence}. This fixed point has many unexpected properties, such as a dynamical critical exponent $z=1$, emergent nesting at the hot spots, and a singular, anisotropic boson dispersion in the infrared.

Finally, it is possible that near the critical point, the system becomes unstable to the formation of some kind of ordered state (different from the primary order that onsets at the QCP). In the presence of time reversal or inversion symmetry, it is natural to expect a superconducting instability--the order parameter fluctuations induce an effective attractive interaction between the fermions, which is enhanced upon approach to the critical point~\cite{Miyake1986,Scalapino1986,monthoux1991toward,Abanov2001quantum,metlitski2010instabilities,metlitski2015cooper,wang2016superconductivity}. On the other hand, the strong scattering of the fermions off the fluctuations causes them to lose their coherence; the resulting superconducting $T_c$ is determined by the competition between these two effects. For weak coupling, and sufficiently far away from the nematic QCP,
superconductivity is enhanced in all symmetry channels~\cite{lederer2015enhancement,Maier2014}.
At the QCP, and for sufficiently small $\lambda$, one can estimate the superconducting susceptibility in the regime $\Omega_{\mathrm{NFL}} \ll T \ll \Omega_b$, by solving the Eliashberg equation for the superconducting vertex, using the form (\ref{eq:Pi}) for the bosonic self-energy (but ignoring the fermionic self-energy, which is small in this regime). For the Ising nematic case, this predicts the superconducting susceptibility to diverge at a scale
\begin{equation}
T_{\mathrm{c,QCP}} \sim \frac{\lambda^4}{E_F} \sim \Omega_{\mathrm{NFL}}.
\end{equation}
Thus, there is no parametric separation between the non-Fermi liquid %temperature at which the Fermi liquid description of the fermion breaks down
and the superconducting scales,
%the scale at which superconducting fluctuations become strong;
%to determine the actual superconducting $T_c$ reliably,
and both of these effects have to be taken into account on equal footing.
%Note also the form of $T_{\mathrm{c,QCP}}$ is fundamentally different from the Bardeen-Cooper-Schrieffer (BCS) one; in particular, unlike in BCS theory, $T_{\mathrm{c,QCP}}$ is not exponentially small in the coupling constant $\lambda$. This is because of the dynamic nature of the induced interactions, that have a singular frequency dependence close to the QCP.

Other types of secondary order parameters have been considered. Close to a ferromagnetic QCP in $d=2$, an instability towards incommensurate spin order was proposed~\cite{Chubukov2004}. Near an SDW transition, it was suggested that instabilities towards charge density wave (CDW)~\cite{metlitski2010instabilities,efetov2013pseudogap,Wang2015} or pair density wave (PDW)~\cite{Agterberg2015} orders may occur. %In particular, within a hot spot theory of an SDW QCP with a linearized dispersion for the fermions, there is an exact symmetry that relates CDW order and superconductivity~\cite{metlitski2010instabilities}. %, implying that the susceptibilities of the two types of orders are identical.

In summary, despite the valuable information provided by the perturbative treatments, there is still uncertainty even about the basic properties of metallic QCPs. Clearly, a non-perturbative, controlled solution is highly desirable.
%, both as a benchmark for existing theories and in order to explore the intermediate to strong coupling regimes.
In this article, we review recent advances in numerical QMC simulations of metallic QCPs, describe the picture that emerges from the existing results, and point to some future directions and outstanding puzzles.

\section{Lattice models and quantum Monte Carlo technique}

Arguably the most powerful numerical approach to solve quantum many-body problems is the quantum Monte Carlo (QMC) technique \cite{Gubernatis2016},
which -- despite an exponentially growing Hilbert space -- allows to sample the expectation values of arbitrary observables with {\em polynomial} effort in system size and inverse temperature. Typically, the problem is formulated as a path integral, mapping it to an effective $(d+1)$-dimensional classical problem with Euclidean time playing the role of an additional dimension. The classical problem is then simulated using the traditional Monte Carlo method~\cite{LandauBinder}. %using, for instance, the Metropolis-Rosenbluth algorithm.
Unfortunately, however, for generic models, the ``Boltzmann weight'' of the effective classical problem can turn out negative (or complex). In general terms, this is because the weights of the $(d+1)$-dimensional system still represent quantum amplitudes, not classical probabilities. In the present context, the occurrence of negative weights can in fact be traced back to the sign that arises in fermionic exchange statistics \cite{troyer2005computational}. Models that suffer from this problem can still be simulated using QMC, but the computational complexity now grows {\em exponentially} with the system size and inverse temperature, thus obliterating the core advantage of the Monte Carlo approach.
% Any extrapolation to the thermodynamic, $T\rightarrow 0$ limit then becomes impractical.
 This is the famous ``sign problem'' \cite{Loh1990}. %, which has hindered the numerical exploration of the many-fermion problem in the same way that systems of many interacting bosonic or spin degrees of freedom have been solved over the last two decades.

Fortunately, as we elaborate below, many models of metallic QCPs can be formulated without a sign problem.
%For this, it is important to note that the sign problem is a basis-dependent problem -- for instance, when formulated in its eigenbasis, any given many-body system will exhibit strictly non-negative weights only. However, finding a suitable sign problem-free basis (which not necessarily needs to be the eigenbasis) is in general highly non-trivial, and doing so in generic terms is known to be an NP-hard problem \cite{troyer2005computational}. However,
Over the past years considerable progress has been made in establishing symmetry criteria for sign-free problems \cite{Lang1993,Wu2005,Chandrasekharan2010,Chandrasekharan2014,li2015solving,li2016majorana,wei2016majorana}. In particular, it has been realized \cite{Wu2005,Wang2015b,Wei2017} that the presence of anti-unitary symmetries of the action matrix play a crucial role. For the models of metallic QCPs at hand, %it will again be a manifestation of
%time-reversal symmetry that allows for the formation of
%pseudospin Kramers pairs, which in turn
such an anti-unitary symmetry is present, and enables a sign problem-free formulation of the problem.
This opens the way to large scale, numerically exact simulations of these systems. In this Section, we describe lattice models that realize metallic QCPs, the specific conditions for the lack of a sign problem, and discuss some merits and limitations of the applied numerical technique. %, and what we may hope to learn from numerical solutions of these models.

\subsection{Ising-nematic quantum critical point}

In order to perform QMC simulations, we first need to formulate the problem on a discrete lattice. %, such as a square lattice.
The microscopic model is then designed such that it realizes a quantum phase transition in the presence of itinerant electrons. To the degree that universality holds, the properties of this transition are independent of the particular lattice realization.
%and are broadly applicable (e.g. in the interpretation of experimental results).
%This assumption, of course, needs to be checked and we describe below some ways to test it.

An Ising-nematic transition in a metal was studied using QMC in Refs.~\cite{Schattner2016,Lederer2017}. To do so, these works introduced a microscopic model containing two sets of degrees of freedom: fermions hopping on a two-dimensional square lattice,
%created by the operators $c^{\dagger}_{\r\sigma}$ (where $\sigma = \uparrow,\downarrow$ is the spin index),
and ``pseudospins'' that reside on the bonds.  %of the lattice.
The Hamiltonian is given by
\begin{equation}
H_{nem} = H_{c} + H_{nem} + H_{int},
\label{eq:H_nem}
\end{equation}
where
\begin{eqnarray}
H_{c}    &=& \sum_{\r, \r', \sigma }  [-t_{\r,\r'} - \mu \delta_{\r,\r'}] c^\dagger_{\r\sigma} c_{\r' \sigma}, \nonumber \\
H_{nem} &=& -h \sum_{\langle \r, \r' \rangle }   \mu^x_{\r,\r'} + V \sum_{ \{\r, \r' \}, \{ \r', \r'' \} } \mu^z_{\r,\r'} \mu^z_{\r',\r''},  \nonumber \\
H_{int} &=& \alpha t \sum_{\langle \r, \r' \rangle,\sigma }   \mu^z_{\r,\r'} \left( c^\dagger_{\r\sigma} c_{\r' \sigma} +h.c. \right).
\end{eqnarray}
Here, $c^{\dagger}_{\r,\sigma=\uparrow,\downarrow}$ are fermion creation operators, $t_{\r,\r'}$ are the hopping matrix elements, $\mu$ is the chemical potential, $\mu^{x,y,z}_{\r,\r'}$ are Pauli matrices that act on the pseudospin degree of freedom at the bond connecting sites $\r$ and $\r'$, $h$ is a ``transverse field'' that controls the quantum fluctuations of the pseudospins, and $V>0$ is the strength of an ``antiferromagnetic'' interaction between the nearest neighbor bonds $\{\r,\r^\prime\}$ and $\{\r',\r''\}$. The strength of the fermion-pseudospin coupling
%between the fermions and the pseudospin degrees of freedom
is set by $\alpha$. In the disordered phase, where $\langle \mu^z_{\r,\r\pm \hat x}\rangle=\langle \mu^z_{\r,\r\pm \hat y}\rangle$%$\langle \mu^z_{\r,\r'}\rangle = \mathrm{const.}$,
the system is symmetric under a $C_4$ rotation by $\pi/2$. For sufficiently small $h$, %$V$ (or small $h$),
there is a transition into a phase where the values of $\langle \mu^z_{\r,\r'}\rangle$ on horizontal and vertical bonds become different. This phase spontaneously breaks the lattice rotational symmetry down to $C_2$. %If $\alpha\ne 0$,
In this nematic phase, the dispersions of the fermions in the $x$ and $y$ directions are different. %This is the nematic phase.

% Why is this model sign free?

In order to set up the model (\ref{eq:H_nem}) for a QMC treatment~\cite{Blankenbecler1981,scalettar-1989}, %one can write
the partition function is written as a discrete-time Euclidean path integral\footnote{The results described here are independent of the size of the time step chosen.}.
%One then proceeds according to the Blankenbecler-Scalapino-Sugar algorithm (BSS)~\cite{Blankenbecler1981,scalettar-1989,AssaadBook}: %and performs a Hubbard-Stratonovich decoupling:
For each space-time configuration of the pseudospins, the fermions can be integrated out exactly, since their action is quadratic. The fermionic contribution to the Boltzmann weight is written as a product of {\em two} fermion determinants, ${\det}(M_\uparrow) \times \mathrm{det}(M_\downarrow)$, where $M_{\uparrow,\downarrow}$ are the action matrices for spin up and down electrons. Crucially,
%since both spin species couple identically to the bosonic degree of freedom,
$M_\uparrow = M_\downarrow$ and both matrices are real. Therefore, the effective bosonic action is real and non-negative, and hence it is amenable to QMC simulations without suffering from the sign problem.

\subsection{Spin density wave quantum critical point}

In the case of a magnetic transition, we are not as fortunate; the magnetic order parameter couples differently to spin up and spin down fermions, and hence integrating out the fermions generally does not produce a real, non-negative Boltzmann weight. Nevertheless, one can formulate a lattice model that realizes a metallic SDW transition, and is free of the sign problem.

In field theoretic treatments of an SDW metallic transitions, one usually focuses on the vicinity of the {\em hot spots}, which are pairs of points on the Fermi surface connected by the magnetic ordering wavevector $\Q$.
%, . At these points, fermions can scatter resonantly off the order parameter fluctuations while remaining on the Fermi surface; therefore, the effect of the bosonic fluctuations on the fermions and vice versa are the strongest.
It is generally believed that any universal properties of the transition are captured within a model that includes only a set of ``patches'' of the Fermi surface in the vicinity of the hot spots. One therefore has the freedom to ``deform'' the Fermi surface away from the hot spots, without qualitatively changing the behavior
%of the system
near the QCP. Let us consider a model with two fermionic flavors, $c_{1,2}$, with different dispersions, %each forming its own Fermi surface,
such that $\Q$ connects points on the Fermi surface of band 1 to points on band 2, but does not connect two points on the Fermi surface of the same band [see Fig.~\ref{fig:FermiSurface}(c)]. Then, to capture the coupling of the fermions to the SDW fluctuations near the hot spots, it is enough to consider only the part of the fermion-boson interaction of the form $\vec{\phi} \cdot (c_1^\dagger \vec{\sigma} c_2 + h.c.)$. The action, regularized on a lattice, then has the following form:
\begin{equation}
S_{SDW} = S_c + S_\phi + S_{int},
\label{eq:SDW_lat}
\end{equation}
where
\begin{eqnarray}
S_{c}  &=&  \sum_{\k, \alpha=1,2,  \sigma} \int d\tau \, c^{\dagger}_{\k,\alpha, \sigma} \left(\partial_\tau +  \varepsilon_{\alpha,\k}  \right) c_{\k,\alpha, \sigma}, \nonumber \\
S_{int} &=& \sum_{\r} \int d\tau \, e^{i \Q \cdot \r} \vec{\phi}_\r \cdot c^\dagger_{\r,1} \vec{\sigma} c^{\vphantom{\dagger}}_{\r,2} + h.c.,
\end{eqnarray}
and $S_{\phi}$ is a lattice version of the bosonic part of the action in Eq.~(\ref{eq:S}), and $c^\dagger_{\r,\alpha} = (c^\dagger_{\r,\alpha,\uparrow}, c^\dagger_{\r,\alpha,\downarrow})$. The order parameter $\vec{\phi}_\r$ can be either a one, two, or three component vector, corresponding to an easy axis, easy plane, or isotropic magnetic order parameter. %Note that the band dispersions of the two flavors, $\varepsilon_{\alpha,\k}$, do not have to be particle-hole symmetric.

The effective bosonic action obtained by integrating out the fermions in~(\ref{eq:SDW_lat}) is real and positive semi-definite, and therefore the model is sign-problem free~\cite{Berg2012}. This is due to the fact that %the two flavor structure of the model:
for every configuration of $\vec{\phi_\r}(\tau)$, the fermion action matrix is symmetric under an anti-unitary transformation $\tilde{\mathcal{T}} = i\sigma_y \mathcal{K} U$, where $\mathcal{K}$ denotes complex conjugation, and $U$ is a unitary transformation that changes $c_1 \rightarrow c_1$, $c_2 \rightarrow -c_2$. Note that
$\tilde{\mathcal{T}}^2 = -1$. The existence of such an anti-unitary symmetry is a sufficient condition for the lack of a sign problem~\cite{Wu2005,Wang2015b,Wei2017}.

\subsection{Determinant Quantum Monte Carlo technique}

We treat the models of Eqs.~(\ref{eq:H_nem}, \ref{eq:SDW_lat}) using the standard determinant quantum Monte Carlo method~\cite{AssaadBook}.
Even in the absence of a sign problem, the method is computationally %rather
costly: its complexity scales as $\beta \mathcal{N}^3$, where $\beta = 1/T$
% is the inverse temperature
and $\mathcal{N}$  is the number of lattice sites. %In two spatial dimensions, this translates into $\beta L^6$, where $L$ is the linear system size.
Nevertheless, one can straightforwardly get to system sizes of $24\times 24$ for the nematic case and $16\times 16$ for the SDW case\footnote{The reason for the difference in the maximum accessible $L$ between the nematic and the SDW models is that in the nematic model there is only one fermionic ``orbital'' per site, whereas in the SDW case there are two orbitals.}, and temperatures as low as $T = 0.025$ in units of the hopping %which corresponds to
(about one percent of the Fermi energy). Fortunately, these system sizes and temperatures are sufficient to extract much of the physics of the QCP in both cases.

The details of the numerical implementations have been %extensively discussed in
described in Refs.~\cite{Schattner2016,Schattner2016a}. %Here we mention only
A few technical tricks turned out to be crucial in order to improve the convergence of the
algorithm:
%tricks are vital to obtain results representative of the thermodynamic limit in a manageable amount of time.
First, global updates and parallel tempering schemes
need to be introduced in the vicinity of the QCP, to overcome critical slowing down. Second, convergence to the thermodynamic limit is dramatically accelerated by simulating systems with a small orbital magnetic field that corresponds to a single flux quantum~\cite{Assaad2002}. The magnetic field is opposite for spin up and spin down fermions, such that it does not introduce a sign problem. Recently, several new directions have been proposed to speed up QMC simulations,
% by using algorithms from machine learning,
that may allow access to larger system sizes~\cite{Junwei2017,Liu2017,Xu2017a,Liu2018}.

Once converged, the simulations give numerically exact results, in the sense that they are devoid of any bias, free of systematic errors, and have statistical errors that can be made arbitrarily small by running the simulations longer. Thus, thermodynamic properties (such as the order parameter susceptibility) and imaginary-time correlation functions (such as the fermion Matsubara Green's function) can be calculated %in a controlled manner
to any desired precision.
Unfortunately, since the simulations are performed in imaginary time, one cannot directly access any {\it real-time} correlation functions. Computing real time (or real frequency) quantities requires an analytic continuation of the imaginary time data, which is a numerically unstable process and requires nontrivial additional assumptions. However, much insight into the physics can be gained by analyzing imaginary-time correlation functions directly, as we highlight below.
%In particular, moments of spectral functions over a frequency window of width $\sim2\pi T$ around $\omega=0$ can be computed, providing some information about the low-frequency properties.

\section{Results}
\label{sec:results}

\subsection{Phase diagrams}
\label{sec:phase_diagram}
% description of the phase diagrams

\begin{figure}
\includegraphics[width = \columnwidth]{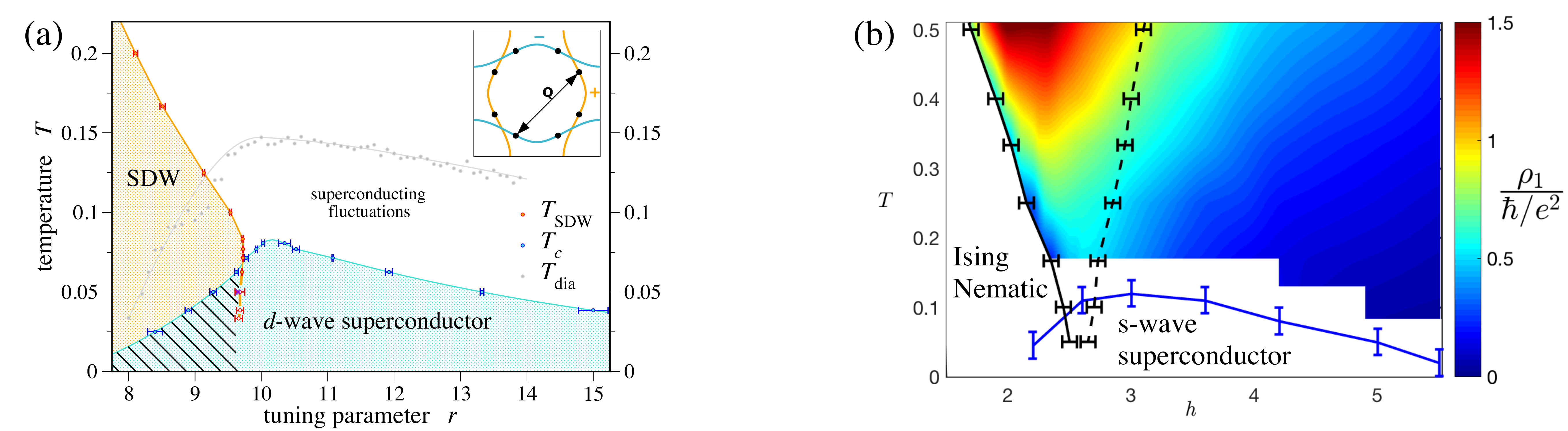}
\caption{(a) Phase diagram of an easy-plane SDW model from Ref.~\cite{Schattner2016}, with a fixed value of the Yukawa coupling, $\lambda=3$. The shaded area marks coexistence between superconductivity and SDW order. The grey line marks the onsets of diamagnetism above $T_c$ (defined as the change of sign of the magnetic susceptibility from paramagnetic to diamagnetic). The inset shows the structure of the superconducting order parameter in $\k$ space. $\Q = (\pi,\pi)$ is the magnetic ordering vector.
(b) Phase diagram of the nematic model [Ref.~\cite{Lederer2017}], as a function of the tuning parameter $h$ and temperature $T$. The dashed lines indicates the the value of $h$ where the nematic susceptibility is half of its value at $h=h_c$ and the same temperature. The color scale shows a proxy $\rho_1$ to the dc resistivity (see Sec.~\ref{sec:transport}). The following parameters were used: $\alpha = 1.5$, $V=0.5 t$, $\mu = t$.}\label{fig:phase_diagram}
\end{figure}

We now review the results of the QMC simulations for the models describing a nematic [Eq.~(\ref{eq:H_nem})] or easy-plane SDW [Eq.~(\ref{eq:SDW_lat})] transition in a metal, studied in Refs.~\cite{Schattner2016a,Lederer2017} and \cite{Schattner2016,gerlach2017quantum}, respectively. The phase diagrams, shown in Fig.~\ref{fig:phase_diagram}, are qualitatively similar: upon increasing the tuning parameter, the ordering temperature\footnote{In the SDW case, the thermal transition has Berezinskii-Kosterlitz-Thouless character, since the order parameter has XY symmetry.} (either nematic or SDW) decreases, extrapolating to zero %indicating the existence of
at a quantum phase transition. In the nematic case, the finite-temperature transition is continuous down to the lowest temperature displayed\footnote{At smaller fermionic densities (not shown) the nematic transition may turn weakly first order.}. In the SDW case, there is some evidence that the transition becomes very weakly first order at the lowest temperature, $T\sim 0.025$, accessible in the numerics.\footnote{Simulations at $T=0$ are possible; however, to the best of our knowledge, none have been performed to date for the easy-plane case.} (All energy scales are in units of the hopping matrix element of the fermions on the lattice.) In both cases, the putative quantum critical point (QCP) is covered by a superconducting phase (detected by measuring either the superconducting susceptibility or the superfluid stiffness), with a maximum $T_c$ occurring near (or slightly to the disordered side of) the QCP. The symmetry of the superconducting order parameter is %similar in the two cases: in the nematic case it is
$s-$wave in the nematic case, whereas in the SDW case it is $d-$wave. For the SDW problem, a regime of substantial superconducting fluctuations between $T_c$ and $\sim2T_c$ is reported  in \cite{Schattner2016}, as manifested by a reduction of the single-particle density of states and an enhanced diamagnetic susceptibility.

% evolution from weaker coupling

%While the phase diagrams in Fig.~\ref{fig:phase_diagram} are calculated for a fixed value of the Yukawa coupling, their dependency on the
%strength of this coupling has been systematically studied in Ref.~\cite{gerlach2017quantum}. For weaker coupling, the phase diagrams
%are found to evolve smoothly with the coupling strength, with the maximum $T_c$ of the superconducting phase decreasing with the coupling strength.

% superconductivity: dependence of Tc on parameters

It is interesting to ask what features of the models control the maximum $T_c$. In the SDW case, this question has been addressed in some detail~\cite{gerlach2017quantum,wang2017superconductivity}. For a fixed band structure, the overall phase diagram evolves smoothly with the strength of the Yukawa coupling $\lambda$, with the maximum superconducting $T_c$ being initially proportional to $\lambda^2$;  %over a range of $\lambda$; then
at larger $\lambda$, $T_c$ saturates to a value of about $0.04E_F$. $T_c$ near the nematic QCP has not been studied at the same level of detail, but seems to behave similarly. Preliminary simulations~\cite{Bauer2018} show that $T_c$ in the isotropic [$O(3)$-symmetric] SDW case has similar trends. In the planar SDW problem, the maximal $T_c$ was found to be strongly dependent on the angle at which the Fermi surfaces meet at the hot spot~\cite{wang2017superconductivity}: over some range of the angle $\theta_{hs}$ and $\lambda$, $T_c \propto \sin(\theta_{hs})$. At larger values of $\lambda$, $T_c$ seems to be less dependent on $\theta_{hs}$.

% masking of the critical point (mention ferromagnetic paper)

In both the Ising-nematic and the planar SDW case, the QCP is inside the superconducting phase. Since the Fermi surface is fully gapped in the superconducting phase in both cases\footnote{Note that the superconducting state in the SDW case posseses a nodeless $d-$wave order parameter, having an opposite sign on the two portions of Fermi surface that are related to each other by $\pi/2$ rotation. See inset to Fig.~\ref{fig:phase_diagram}.}, we expect the QCP to be of the $d=2+1$ Ising or XY types, respectively. Hence, strinctly speaking, a ``pristine'' metallic QCP does not exist in either case.
%
%It is possible that
A metallic QCP may be stabilized by breaking both time-reversal and inversion symmetries, or by applying a magnetic field.
%, which has been a preferred experimental route to suppress superconductivity.
However, in our numerical  QMC simulations, doing so would immediately introduce a sign problem, and this option has not been explored. A model undergoing an Ising ferromagnetic transition~\cite{xu2017non} has shown no superconductivity down to the lowest temperatures considered. However, as the Yukawa coupling is increased, the superconducting correlations grow rapidly, indicating that the ground state is probably superconducting.

Finally, it is interesting to note that besides the primary order parameter (either planar SDW or nematic) and superconductivity, we do not find any substantially enhanced fluctuations of any other form of order, anywhere in the phase diagram. Specifically, we have computed charge density wave (CDW) and pair density wave (PDW) correlations. The superconducting susceptibility is always peaked at $\Q=0$, with no secondary peaks at non-zero $\Q$, indicating no proximate PDW instability. The CDW susceptibility is very moderately enhanced upon approaching the SDW QCP, and then strongly suppressed upon entering the superconducting phase. If particle-hole symmetry is present, the SDW model has an increased symmetry that relates the SC and CDW order parameters,
%implying that
and the susceptibilities of the two order parameters are identical~\cite{metlitski2010instabilities,efetov2013pseudogap}. However, a recent QMC study found that
even a small breaking of particle-hole symmetry strongly lifts the degeneracy between the two types of order in favor of superconductivity~\cite{wang2017charge}.

\subsection{Order parameter correlations}
\label{sec:correlations}
Next, we examine the correlations of the order parameter in the vicinity of the QCP, focusing on the metallic regime above $T_c$.

\begin{figure}
\includegraphics[width = \columnwidth]{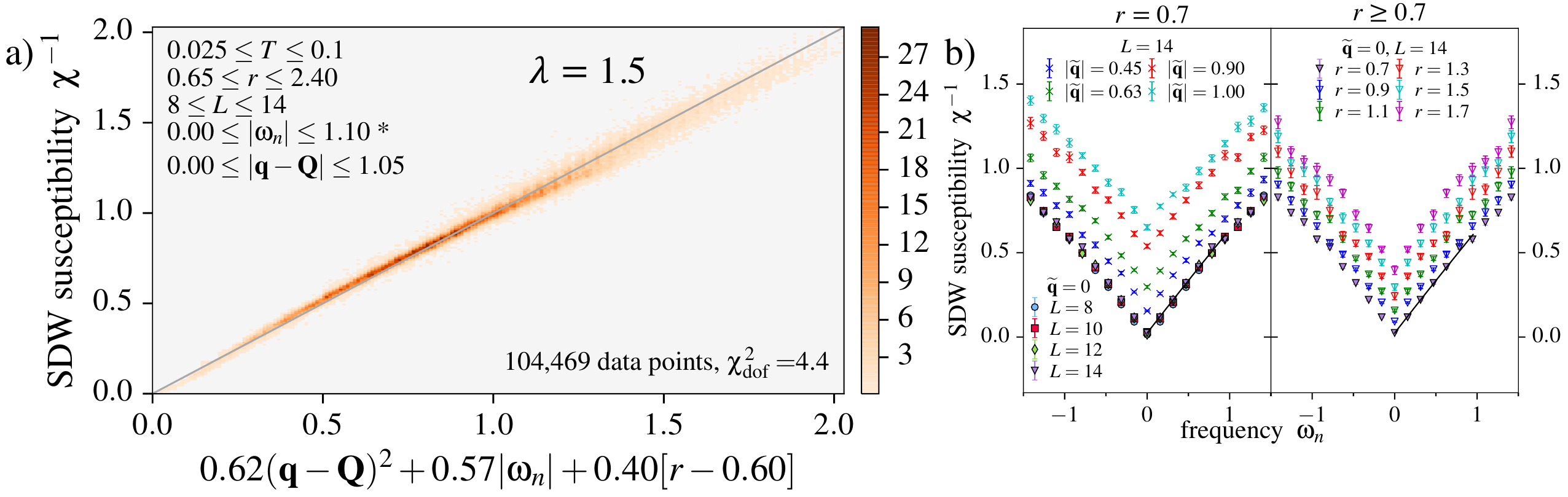}
\caption{(a) The inverse SDW susceptibility, $\chi^{-1}$ near the QCP for different values of $L$, $T$, $\q-\Q$, $\omega_n$, and $r$. The horizontal axis is the functional form $\chi_{\mathrm{fit}}^{-1}$~(\ref{eq:chi_fit}). The color indicates the density of data points for this value of $\chi_{\mathrm{fit}}$. (b) A slice through the same data as in a), showing $\chi^{-1}$ vs. $\omega_n$ for different values of $\tilde{\q}=\q - \Q$ and $r$. In both a) and b), $\lambda=1.5$ and $T=0.025$.}\label{fig:SDW_correlations}
\end{figure}

Fig.~\ref{fig:SDW_correlations} shows the susceptibility, $\chi$, of the planar SDW order parameter, as a function of temperature, tuning parameter, frequency, momentum, and system size. The value of the Yukawa coupling, $\lambda=1.5$, was chosen such that the lowest temperature displayed, $T=0.025$, is just above the superconducting $T_c$. Panel (a) shows that $\chi^{-1}$ can be remarkably well approximated by a simple functional form over the entire regime: $\chi^{-1} \approx \chi_{\mathrm{fit}}^{-1}$, where
\begin{equation}
\chi^{-1}_{\mathrm{fit}} = a |\q - \Q|^2 + b|\omega_n| + c(r - r_c).
\label{eq:chi_fit}
\end{equation}
The values of $a$, $b$, $c$, and $r_c$ are shown in Fig.~\ref{fig:SDW_correlations}(a). Panel (b) shows the dependence of $\chi^{-1}$ on $\omega_n$ for different values of $\q$ and $r$, at $T=0.025$. For other values of $\lambda$, $\chi$ is similarly well described by $\chi^{-1}_{\mathrm{fit}}$.

%As we shall discuss in more detail below,
Eq.~(\ref{eq:chi_fit}) is precisely the form we expect from a naively %approach
integrating out the fermions and treating the resulting effective action at the Gaussian level, {\it \`a la} Hertz-Millis theory~\cite{Hertz1976,Millis1993}. There is one notable difference, however: this treatment~\cite{Millis1993} gives that the ``thermal mass'' [i.e. a temperature dependent additive constant to Eq.~(\ref{eq:chi_fit})] should scale as $T\log(1/T)$. However, in Eq.~(\ref{eq:chi_fit}), we did not include a thermal mass term at all, since it hardly affects the quality of the fit within this resolution. A more careful analysis~\cite{gerlach2017quantum} shows that the thermal mass term is nearly {\it quadratic} in $T$ for $T<0.3$, and nearly linear at higher temperatures.

\begin{figure}
\begin{center}
\includegraphics[width = \columnwidth]{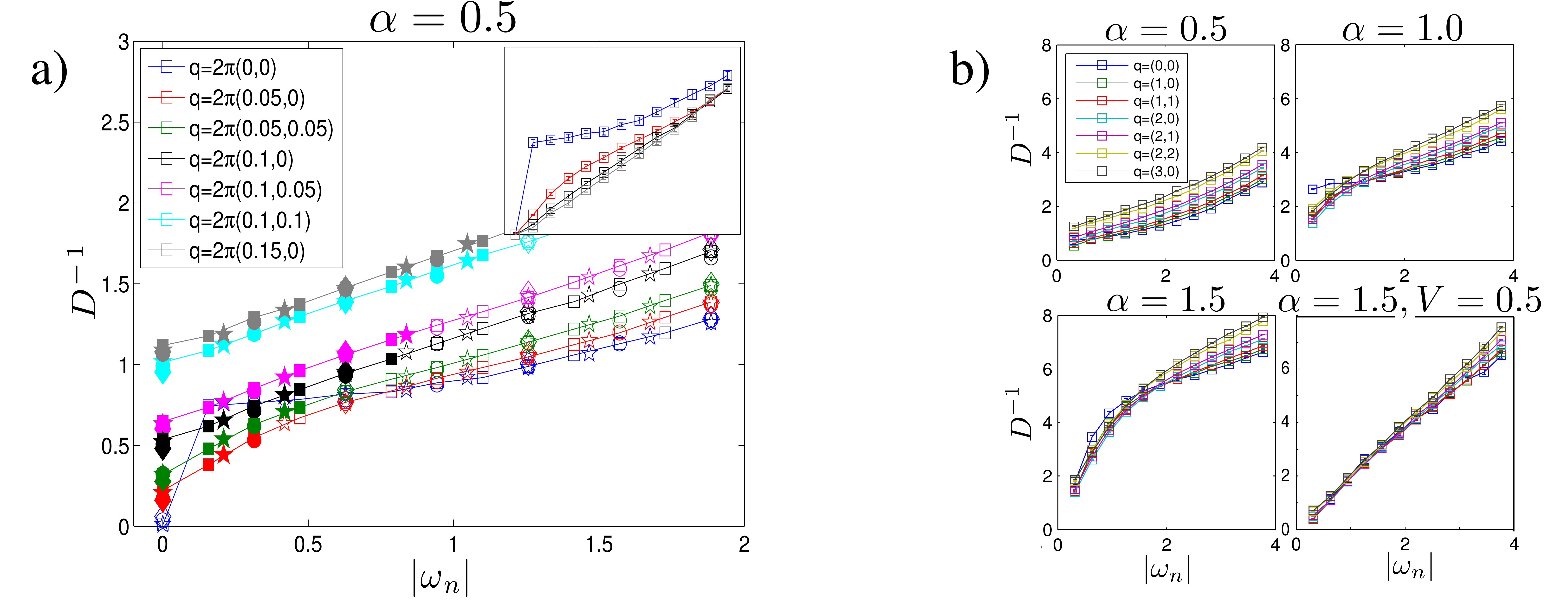}
\end{center}
\caption{The inverse bosonic susceptibility, $D^{-1}$ near the nematic QCP for different values of the Yukawa coupling $\alpha$ [Eq.~(\ref{eq:H_nem})]. a) $D^{-1}$ vs. $\omega_n$ for $\alpha=0.5$ and different values of $\q$ [from Ref.~\cite{Schattner2016}]. Here, $h\approx h_c$, different symbols represent different temperatures: squares, stars, circles, and diamonds correspond to $T=0.025,0.033, 0.05$, and $0.1$, respectively. Solid (open) symbols correspond to points where $v_F|\q| \ge \omega_n$ ($v_F|\q|<\omega_n$), where $v_F$ is the minimal value of the bare Fermi velocity on the Fermi surface. Inset: the same data with the value of $D^{-1}(\omega_n=0)$ subtracted. b) $D^{-1}$ vs. $\omega_n$ for different values of $\alpha$ at $h\approx h_c,\beta=20$. In all the simulations, $V=1,\mu=-0.5$, except the lower right panel in b), where $V=0.5,\mu=-1$. The results are for systems of size $L=20$.}\label{fig:nematic_correlations}
\end{figure}

Next we turn to the order parameter susceptibility near the nematic QCP. We first focus on a moderately small value of the Yukawa coupling, $\alpha=0.5$ in Eq.~(\ref{eq:H_nem}). The static correlations, $D^{-1}(\omega_n=0)$, are well-described~\cite{Schattner2016} by a simple Curie-Weiss form:
\begin{equation}
\tilde{D}^{-1} = A[T + b|\q|^2 + c(h-h_c)] \,.
\end{equation}
The dynamical correlations turn out to be more intricate, and do not permit such a simple description.  Fig.~\ref{fig:nematic_correlations}(a) shows the inverse bosonic propagator $D^{-1}$ as a function of Matsubara frequency for different values of $\q$ in the vicinity of the QCP.
%An apparent feature seen in the data is that
$D^{-1}(\omega_n,\q=0)$ is markedly different from $D^{-1}(\omega_n,\q \ne 0)$, as one may expect for a $\q=0$ order parameter from Hertz theory~\cite{Hertz1976}. $D^{-1}(\omega_n, \q\ne0)$ has an apparent ``cusp'' at $\omega_n=0$, as expected for a bosonic mode Landau damped by the fermions.  %near the Fermi surface.
However, we could not fit $D^{-1}$ to a simple scaling form. As seen in the inset of Fig.~\ref{fig:nematic_correlations}(a), the slope of $D^{-1}$ near $\q=0$ increases with decreasing $|\q|$. The $q$ dependence of this slope is in the same direction as predicted by $|\omega_n|/|\q|$ Landau damping term, but much weaker. This may be due to finite size and finite temperature effects~\cite{Punk2016,Klein2017}. A qualitatively similar behavior was found near an Ising ferromagnetic transition~\cite{xu2017non}, where the ferromagnetic propagator was fit to a modified Hertz-Millis form

%SL I think the analysis leading to this tiny anomalous dimension is shoddy and we shouldn't talk about it \footnote{The $\q$ dependence of the ferromagnetic correlator in Ref.~\cite{xu2017non} is consistent with a small but non-zero anomalous exponent $\eta$; such behavior is not found in the nematic case.}.

At larger values of the Yukawa coupling, $D^{-1}$ shows qualitatively different behavior. The contrast between $D^{-1}(\omega_n=0)$ and $D^{-1}(\omega_n\ne 0)$ becomes less pronounced as the Yukawa coupling increases. Most dramatically, for $\alpha=1.5$, $D^{-1}$ loses its $\q$ dependence almost completely; this is most pronounced when the value of the ``exchange coupling'' between the bosonic degrees of freedom is decreased from $V=1$ to $V=0.5$ [lower right corner of Fig.~\ref{fig:nematic_correlations}(b)]. %SL there is not much evidence for this behavior above T_C.
%This behavior persists all the way down to temperatures of the order of the superconducting $T_c$. %We will discuss its possible origin and consequences in Sec.~\ref{sec:discussion}.

\subsection{Single fermion properties}
\label{sec:spectral}

Near a metallic QCP, we expect the scattering of fermions off the order parameter fluctuations to lead to a breakdown of Fermi liquid behavior. %form of the single-particle Green's function.
We will now address the single-particle properties upon approaching the SDW and Ising-nematic QCPs.

 \begin{figure}
\includegraphics[width=1\columnwidth]{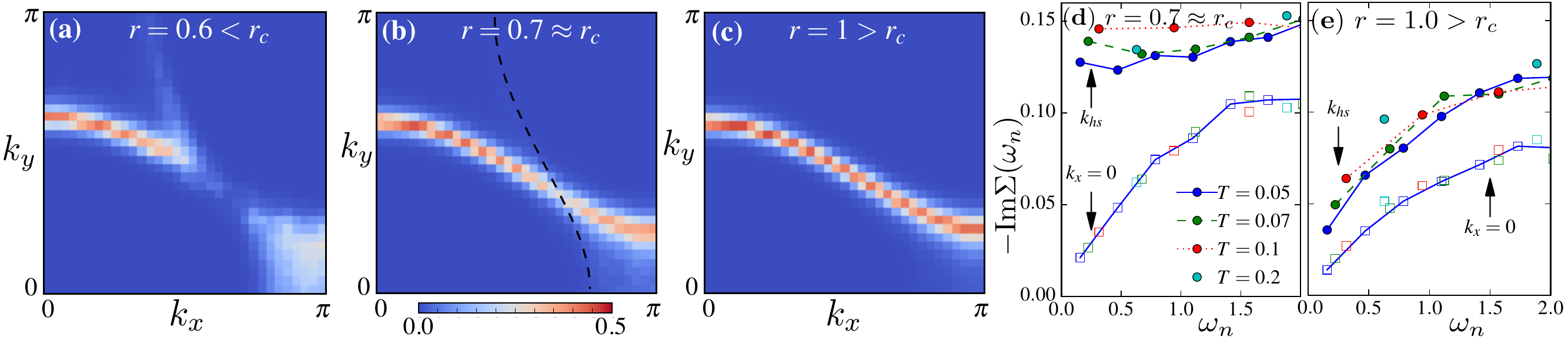}
\caption{Single-particle properties near an SDW QCP with $\lambda=1.5$ [from Ref.~\cite{gerlach2017quantum}]. (a-c) The Green's function $G(\k,\tau=\beta/2)$ as a function of $\k$, for different values of the tuning parameter $r$. (d,e) the imaginary part of the self-energy, $\mathrm{Im} \Sigma(\k_F,\omega_n)$, near the QCP (d) and away from it (e). The self-energy is shown at one of the hot spots ($\k = \k_{hs}$) and at a Fermi surface point with $k_x=0$, away from the hot spots.}
\label{fig:AFM_spectral}
\end{figure}

The imaginary time or Matsubara frequency properties can be computed directly from the QMC simulations. In addition, some information about the properties of the fermion spectral function at real frequencies of the order of $T$ can also be obtained. This is possible through the relation, valid for $0<\tau<\beta$~\cite{Trivedi1995},
\begin{equation}
G(\k,\tau) = \int d\omega\, \frac{e^{-\omega (\tau - \beta/2) }}{2\cosh(\beta \omega/2)} A(\k, \omega),
\label{eq:G_beta_over_2}
\end{equation}
where $A(\k, \omega)$ is the spectral function. From here, we see that $G(\k,\tau = \beta/2)$ gives the integrated spectral weight in a window of width $\sim T$ around the Fermi level. Fig.~\ref{fig:AFM_spectral}(a-c) show a colormap of this quantity vs. $\k$ for one of the orbitals that forms the horizontal Fermi surface shown in Fig.~\ref{fig:FermiSurface}(c), for different values of the tuning parameter $r$ approaching an SDW QCP~\cite{gerlach2017quantum}.  In the ordered state, $r<r_c$ (panel a), the reconstruction of the Fermi surface is clearly visible, and a gap opens at the hot spots. Near the QCP (panel b), the gap at the hot spots fills in, although $G(\k,\beta/2)$ is still significantly suppressed at the hot spots compared to other regions of the Fermi surface. Finally, away from the QCP (c), a full Fermi surface is recovered. %(Note that for non-interacting fermions, $G(\k_F, \beta/2) = 1/2$.)

Next, we examine the fermion self-energy, $\Sigma(\k,\omega_n)$, at different points on the Fermi surface. The imaginary part of $\Sigma(\k,\omega_n)$ is shown at the intersection of the Fermi surface and the $y$ axis (which is far away from the hot spots), and at the hot spot ($\k_{hs}$), either near the QCP or away from the QCP [Figs.~\ref{fig:AFM_spectral}(d,e), respectively.] Away from the hot spots, $\Sigma(\k,\omega_n)$ tends linearly towards zero, consistently with Fermi liquid behavior. At the hot spots, the self-energy is larger than away from the hot spots; this is particularly pronounced in the vicinity of the QCP, where $\Sigma(\k_{hs}, \omega_n)$ is nearly frequency and temperature independent. This marks a strong deviation from Fermi liquid behavior at the hot spots.

The near-independence of the self-energy of $\omega_n$ and $T$ is surprising; most field-theoretical models predict $\Sigma(\omega_n)\sim i \mathrm{sgn}(\omega_n) \sqrt{|\omega_n|}$ at the hot spots. It is not clear whether the behavior found in the simulations represents a true asymptotic property of the QCP.
%Note, however, that
The lowest temperature in the simulations is intrinsically limited by the superconducting transition ($T_c\approx 0.025$ for
%the value of the Yukawa coupling used in this simulation,
$\lambda=1.5$).
Note that for all but the smallest temperature and frequency, $T=0.05$ and $\omega_n = \pi/T$, the self-energy is smaller in magnitude than $\omega_n$, we can then identify a non-Fermi liquid scale $\Omega_{\mathrm{NFL}} \sim 0.05 \approx 2T_c$.
Thus, there is no separation of scales
%As we will further discuss in Sec.~\ref{sec:discussion}, we cannot establish
 between  non-Fermi liquid behavior and superconductivity.

\begin{figure}
\includegraphics[width=\columnwidth]{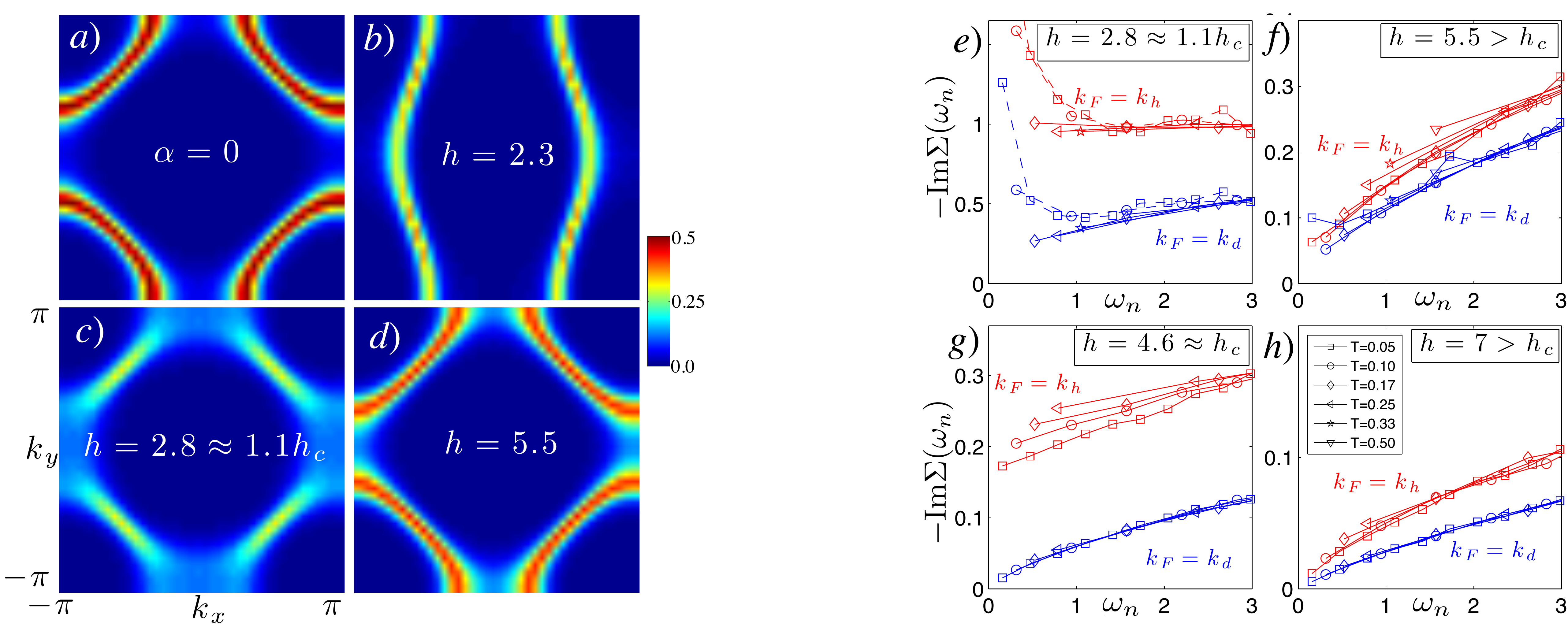}
\caption{Single-fermion properties near a nematic QCP. (a) The imaginary time Green's function $G(\k,\tau=\beta/2)$ as a function of $\k$ for non-interacting electrons ($\alpha$=0) and temperature $T=0.17$. (b-d) $G(\k,\tau=\beta/2)$ for an interacting system with $\alpha=1.5$, $V=0.5$, $\mu=1$, and $T=0.17$ for different values of the tuning parameter, $h$. In panel (b), the system is in the nematic phase; a small symmetry-breaking field was applied in order to pin one of the two configurations of the order parameter. (e,f) Imaginary part of the Matsubara self-energy at the Fermi surface, $-\mathrm{Im}{\k_F, \Sigma(\k,\omega)}$ for two values of $h$, for the same parameters as in panels (b-d). The self-energy is shown at the cold spot on the diagonal, $\k_F = \k_d \parallel (1,1)$, and at a point in the hot region, $k_h$, which is the intersection of the Fermi surface with the segment from $(0,\pi)$ to $(\pi,\pi)$. Similar data are shown in panels (g,h) for a weaker coupling system with $\alpha=1.0$, $V=1.0$, $\mu=0.5$.}
\label{fig:nematic_spectral}
\end{figure}

Next, we describe the single-fermion properties in the nematic case. Figs.~\ref{fig:nematic_spectral}(a-d) show a colormap of $G(\k,\beta/2)$ for different parameters. In the non-interacting case ($\alpha=0$), this function is strongly peaked at the Fermi surface. %, reaching a maximum of $G(\k_F,\beta/2) = 1/2$.
In the presence of interactions ($\alpha=1.5, V=0.5$) but away from the QCP ($h=5.5$), a clear Fermi surface is still present, although some broadening is visible near the ``hot regions'' around $\k=(\pi,0)$ and $(0,\pi)$. In the vicinity of the QCP ($h = 2.8 \approx 1.1 h_c$), a dramatic reorganization of the low-energy spectral weight occurs. While the Fermi surface near the cold spots along the diagonal remains reasonably well defined (even though it is substantially broadened), the Fermi surface in the hot regions is almost completely washed out, with spectral weight distributed over a broad momentum range.

This dichotomy between the cold spots and the hot regions is also apparent in the behavior of the fermionic self-energy, shown in Fig.~\ref{fig:nematic_spectral}(e-h). Panels (e,f) show $\mathrm{Im}\Sigma(\k_F, \omega_n)$ for the parameters used in Fig.~\ref{fig:nematic_spectral}(a) ($\alpha=1.5$, $V=0.5$, $\mu=1$); panels (g,h) show the same quantity for a smaller value of the Yukawa coupling, $\alpha=1$, $V=1$, and $\mu=0.5$. Away from the QCP, both $\mathrm{Im}\Sigma(\k, \omega_n)$ in the hot region [$\k = \k_h$, which is along the line from $(0,\pi)$ to $(\pi,\pi$)] and at the cold spot ($\k = \k_d$, along the diagonal) depend approximately linearly on $\omega_n$.\footnote{The small non-zero extrapolation of $\mathrm{Im}\Sigma(\omega_n)$ to $\omega_n=0$ in Fig.~\ref{fig:nematic_spectral}(f) is likely to be a finite temperature effect; it decreases with decreasing temperature.} In contrast, near the QCP, $\mathrm{Im}{\Sigma}(\omega_n)$ in the hot regions is much larger and does not seem to vanish as $\omega_n\rightarrow 0$, $T \rightarrow 0$. The strong upturn of $\mathrm{Im}{\Sigma}(\omega_n)$ at low frequencies in panel (e) is due to the onset of a superconducting gap\footnote{Compare this behavior to that of the self-energy in the superconducting state within BCS theory, $\Sigma_{\mathrm{BCS}}(\k_F, \omega_n) \sim |\Delta|^2/(i\omega_n)$, where $\Delta$ is the gap.}. At higher frequencies, $\mathrm{Im}{\Sigma}(\k_h, \omega_n)$ in panel (e) is strikingly frequency and temperature independent, reminiscent of the behavior found in the SDW case at the hot spot [Fig.~\ref{fig:AFM_spectral}(d)]. Note that in the stronger coupling simulation of panel (e), $\mathrm{Im} \Sigma(\k_d ,\omega_n)$ has a substantial $\omega_n\rightarrow 0$ extrapolation even at the cold spot; in contrast, in panel $(g)$ it seems to extrapolate to zero.

Thus, both in the SDW and nematic QCPs, strong deviations from Fermi liquid behavior are found at temperatures $T\geq T_c$, at least in some regions of the Fermi surface. The character of these non-Fermi liquid regimes are different from those predicted by theories of metallic criticality; in particular, the self-energy found in the numerical simulations is much less temperature and frequency dependent than expected. The non-Fermi liquid regimes terminate at $T = T_c$, where a gap opens on the entire Fermi surface.

 \subsection{Transport}
 \label{sec:transport}

%Experimentally, the study of transport properties of strongly correlated metals has been particularly insightful in revealing a number of anomalous features, most prominently a linear temperature dependence of the resistivity, that defines an extended ``strange metal" regime.
%Capturing such non-Fermi liquid physics in QMC simulations by probing, for instance, the dependence of the optical conductivity on frequency and temperature,
The frequency-dependent conductivity near a metallic QCP is of particular interest. Unfortunately, computing this quantity from QMC is a highly non-trivial task. Since all dynamical properties are defined in real frequency, they are not easily accessible in QMC simulations performed in {\em imaginary} time. In principle, there is a one-to-one correspondence between the correlation functions in real and imaginary times. %, so one may hope to compute the real-time response from the QMC data.
However, such an analytical continuation involves inverting a nearly singular matrix (with exponentially small eigenvalues).
%-- a particular instance of the Hausdorff problem and as such an ill-conditioned problem.
Hence, any uncertainty in the QMC data (inevitably present because of statistical errors) gets strongly amplified when one attempts to convert the data into real time.

The reverse procedure, of going from real-time spectral functions to imaginary-time correlations, does not suffer from this problem. Therefore, the QMC data can be used to constrain the behavior of the real-time spectral functions, but cannot determine them uniquely. If additional assumptions are made, one can extract model spectral functions that are consistent with the data. For example, the maximum entropy method~\cite{gubernatis1991quantum} attempts to find the least structured spectral function that can reproduce the imaginary-time data.

 \begin{figure}
\includegraphics[width=0.9\columnwidth]{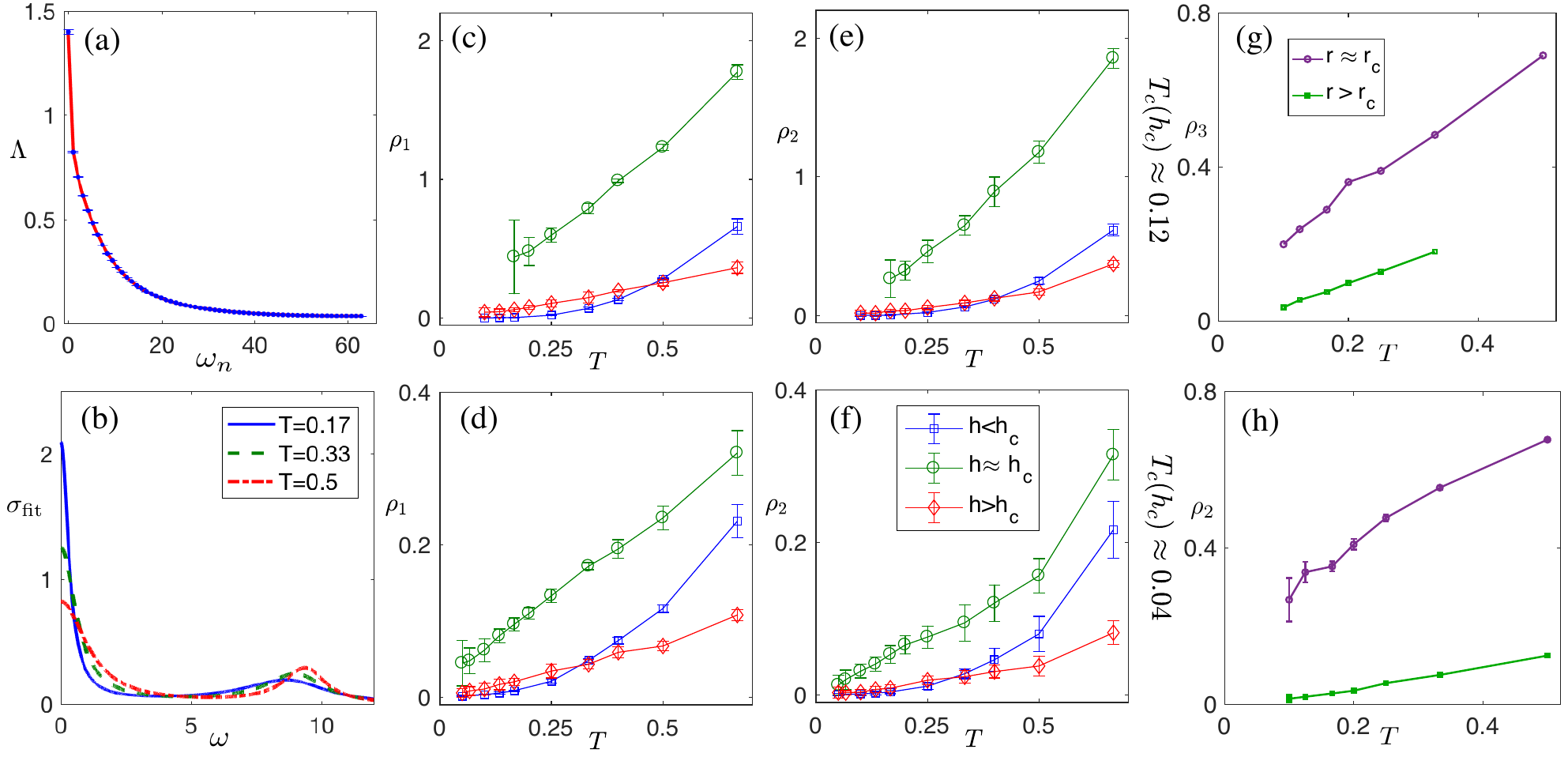}
\caption{(a) Current-current correlator $\Lambda(\omega_n)$ near a nematic QCP~\cite{Lederer2017}. In this simulation, $T=0.17$, $h\approx 2.6$, and the following parameters were used: $\alpha=1.5$, $V=0.5$, $\mu=1$. The red line shows a fit to Eq.~(\ref{eq:Lambda_fit}).
%Inset: fit parameters $\gamma$ and $D=A \gamma$ for the narrow component in the fit, that corresponds to the low-frequency Drude-like peak, as a function of $T$.
(b) Fitted optical conductivity $\sigma_{\mathrm{fit}}(\omega)$ [Eq.~(\ref{eq:sigma_fit})] in units of $e^2/\hbar$, for three temperatures. (c--f) Temperature dependence of the resistivity proxies $\rho_1$, $\rho_2$ (in units of $\hbar/e^2$) for parameters $\alpha=1.5,V=0.5, \mu=1$ [(c,e), for
$h=2.0<h_c$, $h=2.6\approx h_c$ and $h=5.5>h_c$] and $\alpha=1.0, V=1.0,\mu=0.5$
[(d,f), for $h=4.0<h_c$, $h=4.6\approx h_c$ and $h=7.0>h_c$]. For $h<h_c$, a small symmetry breaking field has been applied, such that the resistivity tensor is anisotropic. Only the smaller component of the resistivity proxy is shown. (g,h) Resistivity proxies vs. $T$ near an SDW QCP. Here we set $\lambda=3$, as in Fig.~\ref{fig:phase_diagram}. (g) The maximum entropy result $\rho_3$, and (h) the proxy $\rho_2$, shown near the QCP ($r=10.2$) and deep in the disordered side ($r=20$).}
\label{fig:rho_T}
\end{figure}

Below, we describe the results for the imaginary time current-current correlation function, and several methods that were used to analyze this data and extract information about the real-time optical conductivity, $\sigma(\omega)$. The analysis is similar to the maximum entropy method, and gives similar results; however, it makes the physical assumptions more explicit. %and allows to test their validity.

The Matsubara frequency current-current correlation function $\Lambda(\omega_n)$ is related to the real part of the optical conductivity by
\begin{equation}
\Lambda(\omega_n) = \int \frac{d\omega}{\pi} \frac{\omega^2 \sigma(\omega)}{\omega_n^2 + \omega^2} \,.
\label{eq:Lambda}
\end{equation}
The QMC data for $\Lambda(\omega_n)$ was found to be well-described by the following simple ansatz~\cite{Lederer2017}:
\begin{equation}
\Lambda_{\mathrm{fit}}(\omega_n) = \sum_{j=1}^n\frac{A_j}{\omega_n^2 + \gamma_j |\omega_n| + \Omega_j^2} \,.
\label{eq:Lambda_fit}
\end{equation}
Here, $n$ is the number of ``Lorentz oscillator'' components, and $A_j$, $\Omega_j$, $\gamma_j$ ($j = 1,\dots,n$) are fitting parameters. Analytically continuing this to real frequency, we get the corresponding real part of the conductivity:
\begin{equation}
\sigma_{\mathrm{fit}}(\omega) = \sum_{j = 1}^n \frac{A_j \gamma_j}{(\Omega_j^2 - \omega^2)^2 + \gamma_j^2 \omega^2}.
\label{eq:sigma_fit}
\end{equation}
 It turns out that the minimal number of components required to obtain a good fit to the QMC data is $n=2$: essentially, $\sigma(\omega)$ is composed of a strongly temperature dependent ``Drude-like'' peak, and a broad, weakly temperature dependent background. This procedure yields a ``proxy'' for the dc resistivity, $\rho_1 = 1/\sigma_{\mathrm{fit}}(\omega=0)$. A more conventional maximum entropy analysis yields qualitatively similar results~\cite{Lederer2017}. However, the resulting $\sigma(\omega)$ is not unique: adding more components ($n>2$) gives a fit with a similar quality. This is the main source of possible systematic errors in this analysis, as we discuss further below.

Another useful diagnostic of the low-frequency part of the optical conductivity is obtained by examining the current-current correlation function as a function of imaginary time.
$\tilde{\Lambda}(\tau)$ is related to the conductivity via% the Fourier transform of Eq.~(\ref{eq:Lambda}),
\begin{equation}
\tilde{\Lambda}(\tau) = \int \frac{d\omega}{2\pi} \frac{\omega \cosh[\omega(\beta/2-\tau)]}{\sinh(\omega \beta/2)} \sigma(\omega).
\end{equation}
Thus, we see that derivatives of $\tilde{\Lambda}(\tau)$ at $\tau=\beta/2$ can be interpreted as moments of $\sigma(\omega)$ weighted by a ``window function'' of width $\sim T$:
\begin{equation}
\left[ \left( \frac{d}{d\tau} \right)^{2m}\tilde{\Lambda}(\tau) \right]_{\tau = \beta/2} = \int \frac{d\omega}{2\pi} \frac{\omega^{2m+1} }{\sinh(\omega \beta/2)} \sigma(\omega).
\label{eq:moments}
\end{equation}
If the low-frequency conductivity is characterized by a simple Drude-like peak, Eq.~(\ref{eq:moments}) can be used to extract information about its weight and its width. We define a second resistivity proxy in terms of the moments in Eq.~(\ref{eq:moments}) as
\begin{equation}
\rho_2 = [(d^2 \tilde{\Lambda}/d\tau^2) / (2\pi \tilde{\Lambda})^2]_{\tau = \beta/2}.
\label{eq:proxy}
\end{equation}
If $\sigma(\omega)$ has a Lorentzian form with width much less than $T$, then $\rho_2$ asymptotically coincides with the DC resistivity as $T\to 0$. Moreover, if $\sigma(\omega)$ has a single-peak structure at low frequency {\it and} the width of the peak scales as $T^a$ with $a\le 1$, then $\rho_2$ is proportional to the dc resistivity as $T\to 0$. Using $\rho_2$ as a diagnostic for $\sigma(\omega)$  has the advantage that it does not rely on any fitting procedure. However, it suffers from the same uncertainty as $\rho_1$: it is insensitive to fine structure in $\sigma(\omega)$ on a scale of $\omega \sim T$ or smaller. If such structure exists, then $\rho_2$ does not serve as a good indicator of the dc resistivity.

With these caveats in mind, Fig.~\ref{fig:rho_T} shows the analysis of the transport properties near the QCP, starting with the nematic model. The Matsubara frequency current-current correlation function, $\Lambda(\omega_n)$, near the nematic QCP, is plotted in Fig.~\ref{fig:rho_T}(a). The results of the two-component fit [Eq.~(\ref{eq:Lambda_fit})] for different temperatures at $h\approx h_c$ are shown in Fig.~\ref{fig:rho_T}(b). As the temperature is lowered, the Drude-like component becomes narrower and higher, while its integrated weight is approximately constant.
The corresponding proxy of the dc resistivity, $\rho_1 = 1/\sigma_{\mathrm{fit}}(\omega=0)$, is shown vs. temperature
%for different model parameters
in Figs.~\ref{fig:rho_T}(c,d); it is also shown as a colormap across the $(h,T)$ phase diagram in Fig.~\ref{fig:phase_diagram}(b).

Figs.~\ref{fig:rho_T}(e,f) shows the temperature dependence of the proxy $\rho_2$. It is in qualitative agreement with $\rho_1$: near $h=h_c$, both $\rho_1$ and $\rho_2$ are significantly larger than away from the QCP, and depend nearly linearly on temperature, although for the larger coupling strength ($\alpha=1.5$, upper row) the temperature range of this quasi-linear regime is small. The superconducting phase sets the lower limit on the temperatures we can access. In the strongly coupled case, the resistivity proxies become large at moderate temperatures, exceeding $\hbar / e^2$ at a temperature $T\approx 0.3t \sim 0.1 E_F$.

Similar results are obtained in the SDW model. The data are consistent with an optical conductivity dominated by a Drude-like peak with a width of order $\sim T$ and a nearly constant weight. The maximum entropy analysis result, $\rho_3$, shown in Fig.~\ref{fig:rho_T}(g), is qualitatively consistent with the resistivity proxy $\rho_2$ shown in panel (h).
With the limited temperature range accessible to us, it is difficult to establish a power law behavior for the temperature dependence of $\rho_2$, $\rho_3$. If we assume a power-law behavior $\rho \sim T^x$ at low temperatures, we find $x>1$ for $r>r_c$, and $x<1$ for $r\approx r_c$.

\section{Conclusions and Outlook}
\label{sec:discussion}

The first batch of QMC simulations of metallic QCPs described in this article has yielded remarkable and often unexpected insights,
and opened new questions for numerical and analytical investigation. In this concluding section, we will review and interpret the principal conclusions of these studies and discuss their broader implications.

The most basic fact about both the SDW and Ising-nematic models is that in both cases,
the thermal transitions remain continuous at least down to low temperatures ($T\lesssim 0.01E_F$).
Both models exhibit domes of high temperature superconductivity near their putative QCPs, with a maximum $T_c$ occurring near the QCP, confirming the notion that critical fluctuations are conducive to superconductivity. Notably, in both types of QCPs, superconductivity is the {\em only} contender other than the ``primary'' order parameter. Thus, the physics of ``competing orders'', with many ordered phases in a delicate balance with each other, does not appear to be a ubiquitous feature of metallic QCPs. In addition, no ``pseudogap'' regime is found in the vicinity of the QCPs, in the sense that there is no broad regime characterized by a gap in the fermionic spectrum which is not associated with a nearby ordered phase (such as an SDW or a superconductor).

Consequently, these QCPs are not actually metallic, since they occur within superconducting phases.
Though the models can in principle be supplemented by repulsive interactions and/or symmetry breaking (such as a magnetic field) to suppress superconductivity, the resulting sign problem drastically limits the accessible system sizes and temperatures. %The question remains whether there is an \emph{intermediate asymptotic} quantum critical regime, i.e. whether the crossover scale into the quantum critical regime vanishes parametrically slowly compared to $T_c$ in the weak coupling limit. The presently available QMC data seems to disfavor this possibility. A well-defined critical regime may still exist for smaller coupling strengths than those explored so far. \comment{SL: we need to distinguish non-Fermi liquid from quantum critical here; needs discussing}

Nonetheless, the sign-free models exhibit interesting phenomenology in the region of their quantum critical ``fans" above $T_c$, much of which persists into the superconducting state, except at the lowest frequencies. We can thus operationally call this ``metallic quantum critical behavior''; in this regime, interesting phenomenology is observed in the order parameter correlations, in the fermion Green's function, and in the transport properties.

Some of the robust characteristics of this quantum critical regime, common to both the SDW and Ising-nematic QCPs, are: 1) The order parameter correlations are strongly affected by the coupling to the fermions, and are consistent with overdamped dynamics of the bosons. 2) The single-particle fermionic correlator shows non-Fermi liquid behavior of an unexpected kind. Strikingly, the fermionic self-energy is nearly frequency and temperature independent over a broad regime above the superconducting $T_c$. 3) The transport properties show strong anomalies, consistent with a significant broadening of $\sigma(\omega)$ near the QCP. If the results are interpreted in terms of a single ``Drude-like'' peak, the data are consistent with a large dc resistivity, nearly linear in temperature.
%Below, we elaborate on these results and discuss their consequences.

The order parameter dynamics of the SDW model are in broad agreement with Hertz-Millis theory, with the exception of the temperature dependence of the order parameter susceptibility (the latter displays a crossover behavior from $1/T$ to $1/T^2$ upon decreasing temperature). The dynamical critical exponent is $z=2$.
In the Ising-nematic case, the static order parameter correlations are well-described by a simple mean field-like form, again in agreement with Hertz-Millis theory. However, the order parameter dynamics do not admit such a simple description. While at relatively weak values of the coupling, it is qualitatively similar to the $|\omega_n|/|\mathbf{q}|$ form expected from Hertz-Millis theory, its form evolves substantially as the coupling strength is increased.
At stronger couplings, the $\omega_n\ne 0$ propagator becomes increasingly momentum independent (i.e., the effective dynamical critical exponent increases).
It is tempting to associate this behavior with an emergent ``local quantum criticality'',
although the present studies of the Ising-nematic model do not yet establish generic order parameter dynamics in the strong coupling regime.

The single fermion correlations in the quantum critical regime are simpler than the order parameter correlations, but also more unexpected. At weaker coupling,
the fermionic correlations are consistent with a renormalized Fermi-liquid, while the order parameter correlations are substantially altered, consistent with the conclusions of Sec.~{\ref{sec:perturbative}}. At stronger coupling, in both the SDW and Ising-nematic cases, there is a non-Fermi liquid regime above $T_c$ \footnote{It remains to be seen whether the non-Fermi-liquid regime extends to temperatures parametrically higher than $T_c$ in an appropriate limit.}. This regime exhibits substantial loss of quasiparticle coherence in the ``hot'' regions of the Fermi surface: the hot spots in the SDW case, and everywhere but the cold spots in the Ising-nematic case.
At these momenta, the imaginary part of the self-energy is nearly constant, with only weak dependence on frequency and temperature. This violation of Fermi liquid behavior is stronger than predicted by field theoretical methods, which predict a power law dependence of the self-energy on frequency.%SL with exponents $1/2$ and $2/3$ for the SDW and Ising-nematic cases, respectively.

The origin of this behavior is yet to be clarified. It may well be a characteristic of a finite-temperature crossover, not representative of the asymptotic infrared properties of any QCP. (Note, however, that this behavior extends down to temperatures close to $T_c$.) In our opinion, it is worth considering the more radical possibility that the metallic quantum critical ground state is in fact characterized by a nonzero $\mathrm{Im}(\Sigma)$ in the $\omega_n\to 0$ limit.

The transport signatures of the two models show unambiguously that the optical conductivity is strongly affected by the proximity of the QCP. The implications for the dc transport, however, merit both excitement and caution. The data in both models are consistent with $T$-linear resistivity near the QCP;  however, this analysis is subject to the usual uncertainties associated with analytical continuation of imaginary time data. In particular, it relies heavily on the assumption that there is no structure in $\sigma(\omega)$ at frequencies much smaller than $T$. Developing ways of independently testing the consistency of this assumption within the simulations is desirable.

Future investigations of metallic quantum critical points will hopefully develop both the breadth and depth of our understanding. The robustness of the behavior described in this review should be tested by simulating microscopic models with different band structures and different kinds of broken symmetry. Further direct comparisons of the QMC data to perturbative analytical calculations, at least in the weak to intermediate coupling regimes,
can provide useful insights into the interpretation of the results.
Finally, sign-problem free models, such as the ones often used to describe metallic QCPs, form a ``zero measure'' set in Hamiltonian space. It is important to study to what extent the results presented here carry over in the presence of more generic interactions. Devising ways to do this numerically is, of course, a highly challenging problem at the frontier of the field of many-body quantum physics.

\vskip 5mm

\section*{Acknowledgements}

It is a pleasure to thank C. Bauer, S. Chatterjee, D. Chowdhury, A. Chubokov, R. Fernandes, M. Gerlach, S. Kivelson, A. Klein, Z.-Y. Meng, M. Metlitski, S. Sachdev, K. Sun, X.-Y. Xu, Xiaoyu Wang, and Yuxuan Wang for collaboration on the numerical investigations reviewed here and on related topics. S.L. acknowledges support from the Gordon and Betty Moore Foundation's EPiQS Initiative. Y.S is supported by the Department of Energy, Office of Science, Basic Energy Sciences, Materials Sciences and Engineering Division, under Contract DEAC02-76SF00515 and by the Zuckerman STEM Leadership Program.
We also gratefully acknowledge allocation of the much needed computing time on the CHEOPS cluster at RRZK Cologne, the JURECA cluster at the Forschungszentrum J\"ulich, the SIMES cluster at SLAC, and the ATLAS cluster at the Weizmann Institute.

\bibliographystyle{ar-style4}

\bibliography{References.bib}

\end{document}